\documentclass[11pt,twoside]{article}

\usepackage{asp2004}
\usepackage{epsf}
\usepackage{psfig}
\usepackage{lscape}

\begin{document}
\title{Science Results Enabled by SDSS Astrometric Observations}   
\author{\v{Z}eljko Ivezi\'{c}$^1$, Nicholas Bond$^2$, Mario Juri\'{c}$^2$, Jeffrey A. Munn$^3$, 
Robert H. Lupton$^2$, Jeffrey R. Pier$^3$, Gregory S. Hennessy$^4$, Gillian R. Knapp$^2$, 
James E. Gunn$^2$, Constance M. Rockosi$^5$, Tom Quinn$^1$}   

\affil{$^1$University of Washington, $^2$Princeton University, $^3$USNO Flagstaff, 
       $^4$USNO Washington D.C., $^5$University of California--Santa Cruz}

\begin{abstract} 
We discuss several results made possible by accurate SDSS astrometric measurements
in a large sky area, with emphasis on asteroids and stellar proper motions obtained 
by comparing POSS and SDSS. SDSS has observed over 200,000 moving objects in five 
photometric bands, corresponding to about two orders of magnitude 
increase over previous multi--color surveys. These data were used to extend the measurement
of asteroid size distribution to a smaller size limit, to demonstrate that asteroid 
dynamical families, defined as clusters in orbital parameter space, also strongly 
segregate in color space, and to discover a correlation between asteroid age and colors. 
A preliminary analysis of SDSS-POSS proper motions for $\sim$1 million M dwarf stars
demonstrates that, in the 0.1--1 kpc distance range, the rotational velocity and its 
dispersion for disk stars increase with the distance from the Galactic plane.

\end{abstract}

\section{Introduction}
The astrometric obervations obtained by modern massive digital sky surveys, such as 
SDSS, 2MASS, and FIRST, are enabling studies that have not been possible 
before. In this contribution we describe several science results enabled by SDSS astrometric
observations, with emphasis on observations of asteroids and proper motions obtained by 
comparing POSS and SDSS astrometric measurements.

The Sloan Digital Sky Survey (SDSS) is currently mapping one quarter of the sky in 
five optical bands ({\it ugriz}, Fukugita et al.~1996; Gunn et al.~1998) to a depth 
of $r\sim22.5$, accurate to $0.02$ magnitudes 
(both absolute calibration, and root-mean-square scatter for sources not limited by 
photon statistics; Ivezi\'{c} et al.~2004). Astrometric positions are accurate to better 
than $0.1\arcsec$ per coordinate (rms) for sources with $r<20.5$ (Pier et al.~2003), 
and the morphological information from the images allows reliable star/galaxy separation 
to  $r\sim 21.5$ (Lupton et al.~2002). The survey's coverage of $\sim10^4$~deg$^2$ in the North 
Galactic Cap and of $\sim200$~deg$^2$ in the Southern Galactic Hemisphere will result in
photometric and astrometric measurements for over $10^8$ stars and a similar number of 
galaxies. Additionally, SDSS will obtain spectra for over $10^6$ objects, including $10^6$ 
galaxies and $10^5$ quasars. The recent third public Data Release (DR3) includes imaging 
data for $5282$ deg$^2$ of sky, and catalogs $1.4 \times 10^8$ objects (Abazajian et al.~2005). 
A detailed technical description of SDSS is given by Stoughton et al. (2002).

The SDSS astrometric reductions, and the measurements of their success, are described 
in detail by Pier et al. (2003). Briefly, the SDSS absolute astrometric accuracy is
better than 100 mas, with relative (band-to-band) accuracy of about 20-30 mas (rms, for 
sources not limited by photon statistics). In addition to providing positions for a large 
number of objects with a remarkable accuracy (and thus enabling recalibration of other less 
accurate surveys, as described below), an important characteristic of SDSS astrometric 
observations is that measurements in five photometric bands are obtained over a five minute 
long period (with 54 sec per exposure). The multi-color nature allows the discovery of 
the so-called Color Induced Displacement binary stars (for details see Pourbaix et al. 
2004), and the time delay allows the recognition of moving objects, discussed next.

\section{SDSS Observations of Solar System Objects}

Although primarily designed for observations of extragalactic objects, the SDSS is 
significantly contributing to studies of solar system objects, because asteroids in 
the imaging survey must be explicitly recognized to avoid contamination of the quasar 
samples selected for spectroscopic observations (Lupton {\it et al.} 2002). The SDSS 
has already increased the number of asteroids with accurate five-color photometry by 
about two orders of magnitude (to over 200,000), and to a limit more than five magnitudes 
fainter than previous multi-color surveys (Ivezi\'{c} {\em et al.} 2001).

About 43,000 of those 200,000 objects have been associated with previously known 
asteroids that have well determined orbital elements (Juri\'{c} {\em et al.} 2002). 
Both SDSS data and orbital elements are available from the public SDSS Moving Object 
Catalog\footnote{Available from http://www.sdss.org/dr2/products/value\_added/index.html} 
(Ivezi\'{c} et al. 2002a). The SDSS observations of objects with known orbits 
vividly demonstrate that asteroid dynamical families, defined as clusters in orbital 
parameter space, also strongly segregate in color space (Ivezi\'{c} et al. 2002b).
This segregation indicates that the variations in chemical composition within a family 
are much smaller than the compositional differences between families, and strongly 
support earlier suggestions that asteroids belonging to a particular family have a 
common origin.

Asteroid colors measured by SDSS are well correlated with the family age (Jedicke et al.
2004), which provides a direct evidence for space weathering, and offers a method to date 
asteroids using their SDSS colors. The effects of space weathering can also explain
the color variability of asteroids measured by SDSS (Szabo et al. 2004).

\begin{figure}[t]
\plotfiddle{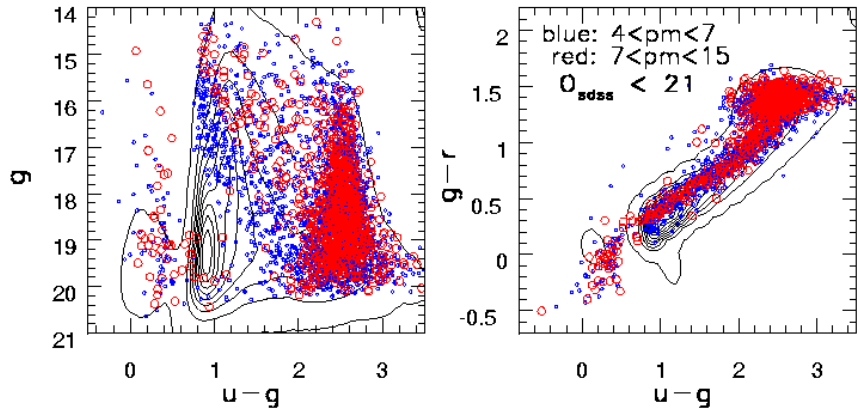}{4.5cm}{0}{60}{60}{-185}{-170}
\caption{An illustration of the differences in the distribution
of point sources in SDSS color-magnitude and color-color diagrams
induced by requiring a large proper motion. The distribution of all
point sources is shown by contours, and two subsets of sources with
large proper motion are shown by symbols (small: 4--7 mas/yr; large: 7--15 
mas/yr). The white dwarfs are easily recognized by their blue $u-g$ color 
($u-g<0.8$), while the majority of other sources with large proper motions 
are main sequence stars. M dwarfs, at distances up to $\sim$1 kpc, dominate 
the large proper motion sample. 
\label{pmcuts}}
\end{figure}

\section{ Proper Motions Determined from SDSS and POSS Observations}

\begin{figure}[t]
\plotfiddle{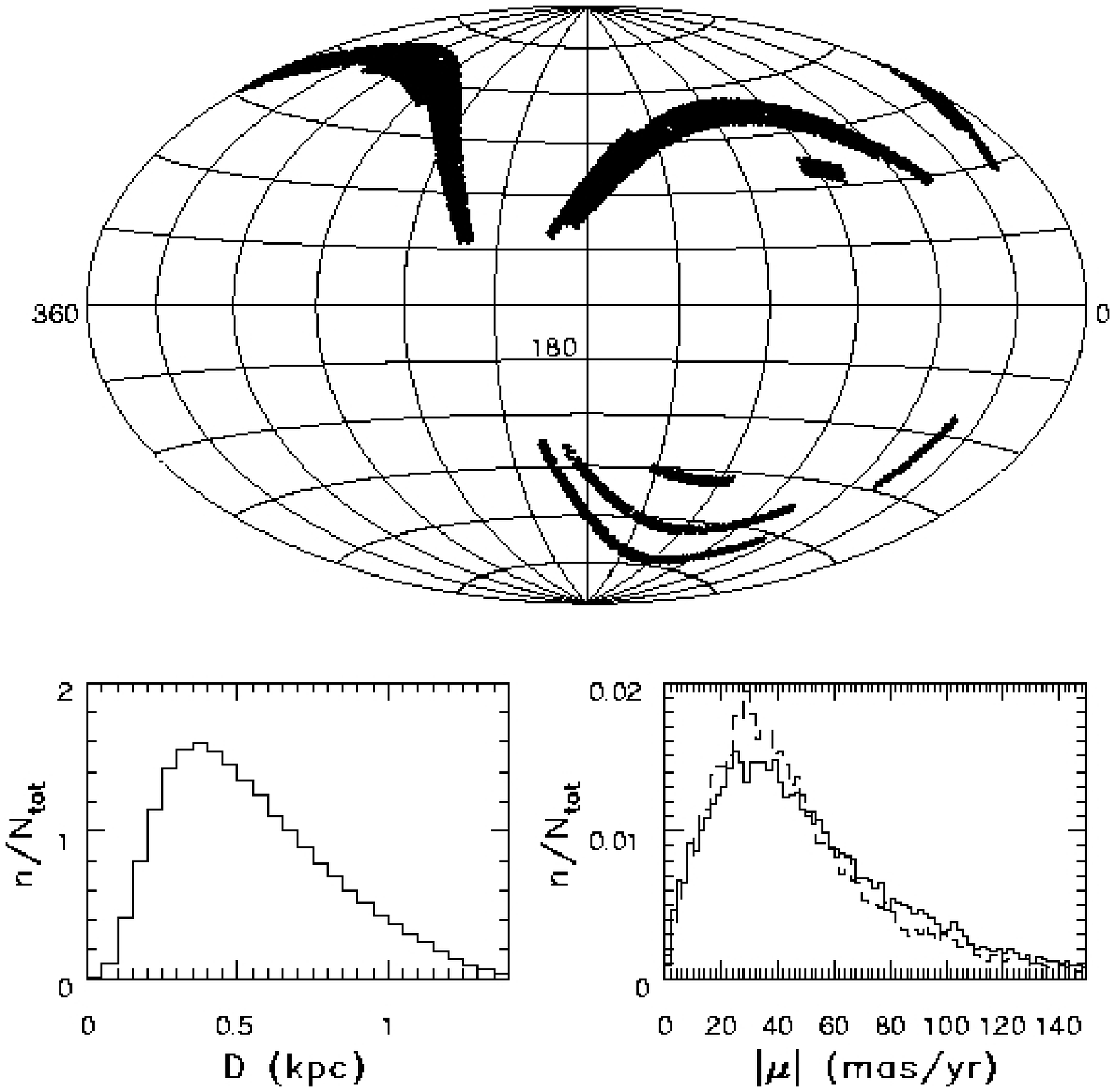}{7cm}{0}{50}{50}{-160}{-117}
\caption{The bottom left panel shows the distance distribution for a sample
of 881,913 M dwarfs with SDSS-POSS proper motions. The top panel shows the sky 
distribution for a subsample with distances in the range 100--150 pc in Aitoff 
projection of galactic coordinates, and the bottom right panel compares the proper 
motion distribution for this subsample before (solid) and after (dashed) correction 
for solar peculiar motion (see \S~\ref{Ss}). 
\label{aitoff}}
\end{figure}

Munn et al. (2004, hereafter M04) have recently presented an improved proper-motion catalog for
$\sim$8 million stars, based on combining the USNO-B and SDSS catalogs in the 2099 deg$^2$ 
large area of sky covered by SDSS Data Release 1. USNO-B positions are recalibrated using 
SDSS galaxies, which results in smaller astrometric errors and the placement of proper 
motions on an absolute reference frame. Tests based on several tens of thousand quasars 
spectroscopically confirmed by SDSS show that the proper motion errors are $\sim$3 mas/yr 
for bright ($g<18$) sources (rms per coordinate), with substantially smaller systematic 
errors ($<$0.5 mas/yr, and typically $\sim$0.1 mas/yr), and excellent stability across 
the sky. For fainter sources, the proper motion 
errors increase to $\sim$4 mas/yr at $g\sim19$, $\sim$5.7 mas/yr at $g\sim20$, and to 
$\sim$7.5 mas/yr at $g\sim20.5$, the faint limit for the sample discussed here.
For stars at 500 pc, these proper motion errors correspond to systematic velocity 
error of $<$1 km/s, and to random velocity errors of 7 km/s at the bright end (without 
including errors in distance), and 20 km/s at the faint end.

This accurate proper motion database offers numerous possibilities for studying the stellar
kinematics, but the analysis of such a large and highly-dimensional data set is not 
trivial. At the very least, the proper motion components are functions of at least
four observables: apparent magnitude, distance (or absolute magnitude, measured by e.g. 
$r-i$ color), and the position on the sky (Fig.~\ref{pmcuts}). Additional stellar parameters, 
such as metallicity (measured from spectra, or estimated from the $u-g$ color), are also 
expected to play a role. Furthermore, the probed distance range is large (out to
$\sim$15 kpc with main sequence stars, Juri\'{c} et al. 2005) and thus the dependence
of the stellar velocity distribution on position in the Galaxy could also be important,
as well as the deviations from a Gaussian distribution.

In this contribution we limit our analysis to M dwarfs from the M04 proper motion catalog 
because 1) they dominate the high proper motion sample, and 2) their distances can be reliably 
determined from a photometric parallax relation. A more comprehensive analysis of that catalog, 
that also includes three-dimensional velocities for $\sim$100,000 stars with SDSS spectra, 
will be presented by Bond et al. (2005, in prep., hereafter B05). 

We select 881,913 M dwarfs from the M04 catalog by requiring $1.25<g-r< 1.50$, $0.6 < r-i < 1.6$ 
and $15 < g < 20.5$ (Ivezi\'{c} et al. 2004), and estimate their distances using 
a photometric parallax relation:
\begin{equation} 
   M_i = 4.0 + 10.86 \,(r-i) -10.74 \, (r-i)^2 + 5.99\, (r-i)^3 - 1.20\, (r-i)^4.
\end{equation} 
These distances have random errors of $\sim$10\%, with comparable systematic errors
(Juri\'{c} et al. 2005). 
The sky and distance distributions of selected stars are shown in Fig.~\ref{aitoff}. 
The size of this sample and its sky distribution make it well suited for
a determination of solar peculiar motion, discussed next.

\begin{figure}[h]
\plotfiddle{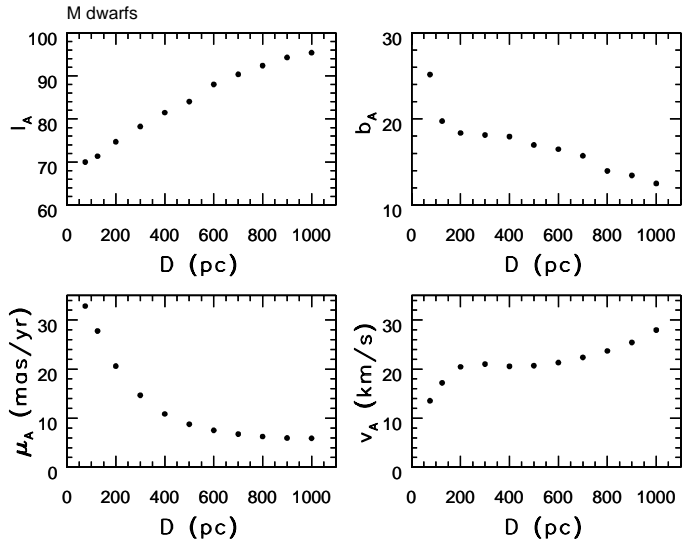}{6cm}{0}{50}{50}{-150}{-70}
\caption{The galactic coordinates of the solar apex (the top two panels),
implied proper motion magnitude (bottom left), and corresponding 
solar velocity (bottom right), determined as a function of distance 
using SDSS-POSS proper motions for a sample of 881,913 M dwarfs. 
\label{apex}}
\end{figure}

\subsection{ Determination of Solar Motion Relative to M dwarfs }
\label{Ss}

The solar peculiar motion with a magnitude of v$_\odot$, induces a 
dipole in the proper motion distribution for sources at a distance $D$, 
with a magnitude of $\sim$3.2\,(v$_\odot$/15 km/s)\,(kpc/D) mas/yr,
which is easily measurable with the sample discussed here. We follow
a procedure outlined by Mihalas \& Binney (1981), and determine the apex
coordinates and the magnitude of implied solar motion for subsamples 
selected in narrow distance bins (Fig.~\ref{apex}). 

The extrapolation of results shown in Fig.~\ref{apex} to zero distance
gives (l$_A$=68$^\circ$, b$_A$=32$^\circ$, and v$_A$=9 km/s), 
somewhat different from standard values (l$_A$=51$^\circ$, b$_A$=23$^\circ$, 
and v$_A$=15 km/s; Mihalas \& Binney 1981). However, as evident 
from Fig.~\ref{apex}, all three quantities vary as a function of the 
subsample distance. We find that this effect is caused by the increase of
the rotational velocity component with the distance from the plane,
which breaks the dipole assumption of the Mihalas \& Binney method. Consequently,
the apex is biased toward (l$\sim$90$^\circ$, b$_A$=0$^\circ$) direction, 
with a corresponding increase of v$_A$. 
An illustration of this dependence is discussed next. 
Additional supporting evidence for this claim, which utilizes three-dimensional 
velocity information, will be described in N05.

\subsection{ Kinematics as a Function of $Z$ }

\begin{figure}[t]
\plotfiddle{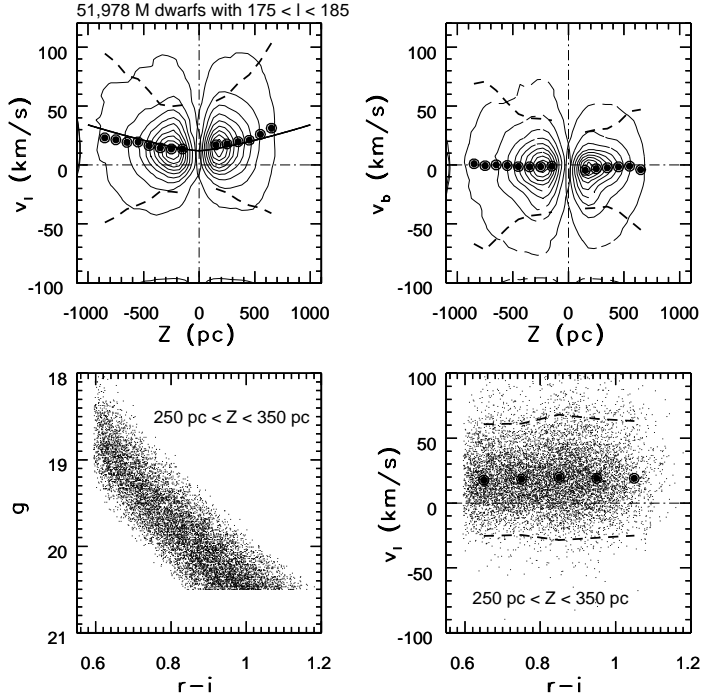}{8.2cm}{0}{50}{50}{-150}{-40}
\caption{The contours in the top left panel show the longitudinal velocity component
determined from proper motion measurements, $v_l$, as a function of distance from the 
Galactic plane, $Z$, for 51,978 M dwarfs with $175 < l <185$. Toward this direction, 
the longitudinal velocity is equivalent to rotational velocity. The large dots are 
the median values of $v_l$ for a given $Z$ bin. The dashed lines represent 2$\sigma$ 
envelope around the median ($\sigma$ is determined using the interquartile range). 
The line following the medians is $v_l = (12 + 22\,|Z/{\rm kpc}|^{1.3})$ km/s. 
The top right panel is analogous, except that the latitudinal velocity component is shown. 
The bottom left panel shows the $g$ vs. $r-i$ color-magnitude diagram for a subsample of 9,231 
stars with $250 < Z/{\rm pc} < 350$. The 
longitudinal velocity component of those stars as a function of $r-i$ color is
shown in the bottom right panel, with the same meaning of symbols and lines as in 
the top two panels. Note that neither the median, nor the distribution width, vary 
with color, although the apparent magnitude, $g$, varies by $\sim$2 mag.
\label{pm1}}
\end{figure}

Here we present results for the kinematics of M dwarfs observed towards the Galactic 
anticenter. In this direction, the rotational velocity is dominated by the 
longitudinal velocity component measured from proper motion, and thus robust 
results can be derived without radial velocity information. The top two panels
in Fig.~\ref{pm1} show the dependence of longitudinal and latitudinal velocity
components, determined from proper motions, for $\sim$50,000 M dwarfs toward
l$\sim$180 (without including correction for solar peculiar motion, which
amounts to an overall shift of the longitudinal component by $\sim$10 km/s). 
While the median values of the latitudinal component are assuringly
consistent with zero, the medians of the longitudinal component increase with
the distance from the plane, $Z$, as v$_l$ = $(12 + 22\,|Z/{\rm kpc}|^{1.3})$ km/s
(with the uncertainty of $\sim$10\% for  all three fitted parameters).

From the data presented here it cannot be ruled out that the observed 
variation of v$_l$ actually reflects a dependence on the radial distance from 
the Galactic center (because the range of galactic latitudes observed by 
SDSS towards l$\sim$180 is limited). However, N05 demonstrate, using three-dimensional
velocity information, that the $Z$ dependence dominates. Furthermore, they
also show that the same relation describes the behavior of v$_l$ for bluer
stars that sample distances out to $\sim$3 kpc (though with less reliable
distances).

\begin{figure}[t]
\plotfiddle{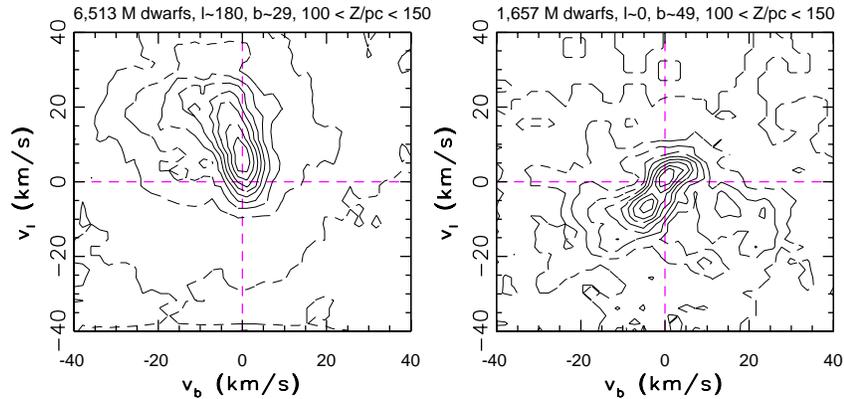}{5cm}{0}{65}{65}{-250}{-470}
\caption{The two-dimensional velocity distribution, $v_l$ vs. $v_b$, determined
from proper motion measurements for two characteristic lines of sight (left: $175 < l < 185$, 
right: $355 < l < 5$) and for the height above the plane in the range 100--150 pc.
The density of sources is shown by linearly spaced isopleths. 
\label{vel2D}}
\end{figure}

\begin{figure}[t]
\plotfiddle{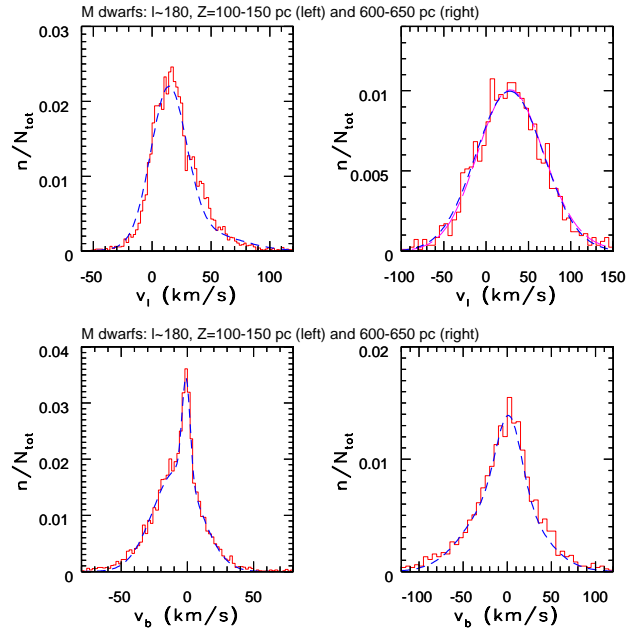}{7cm}{0}{80}{80}{-130}{-390}
\caption{An analysis of the shape of the $v_l$ (top) and $v_b$ (bottom) distributions
toward l$\sim$180, and for two ranges of distance from the Galactic plane (left: 100--150 pc,
corresponding to the left panel in Fig.~\ref{vel2D}, right: 600--650 pc). The histograms 
shown by thin full lines are the data for 7,042 (left) and 1,705 (right) stars, and the 
dashed lines show empirical models. In the top left panel, the model is a sum of two 
Gaussians ($\mu$, $\sigma$) = (14, 15) and (38, 34) km/s, with the wider component accounting 
for 43\% of the sample. In the top right panel, the short-dashed line is a Gaussian with 
($\mu$, $\sigma$) = (28, 40) km/s, and the thin long-dashed line (almost indistinguishable) 
is the convolution of the data shown in the top left panel with a ($\mu$, $\sigma$) = 
(0, 34) km/s Gaussian, and shifted right by 10 km/s. Since the expected measurement errors 
are much smaller (random $\sim$21 km/s, and systematic $<$1 km/s), they cannot explain 
the observed change of the $v_l$ distribution shape as the distance increases from 100 to 
600 pc. In the bottom left panel, the model is a sum of two Gaussians ($\mu$, $\sigma$) 
= (-1, 3) and (-8, 18) km/s, with the wider component accounting for 86\% of the sample. 
In the bottom right panel, the model is a sum of two Gaussians ($\mu$, $\sigma$) 
= (2, 14) and (-5, 38) km/s, with the wider component accounting for 73\% of the sample. 
\label{shapes}}
\end{figure}

\subsection{The Shape of the Velocity Distribution }

The analysis described in preceeding sections considers only the first
two moments of the velocity distribution. Nevertheless, the large number of 
stars allows an unprecedentedly robust and accurate analysis of the full shape 
of the velocity distribution. As Figs.~\ref{vel2D} and \ref{shapes} show, 
the observed shapes significantly deviate from Gaussian, and show distortions 
similar to those expected from the asymmetric drift effect (Mihalas \& Binney 1981). 
The shapes also vary with distance, a variation that is hard to explain
as due to increasing proper motion errors toward the faint end (Fig.~\ref{shapes}). 
The observations of the shape of the velocity distribution, and its dependence on 
the position and stellar tracer, encode an enormous amount of information (e.g. Binney 
\& Tremaine 1987) that we are only beginning to harvest.

\section{Discussion}

The availability of modern massive sky surveys is rapidly changing astronomical
methodology, and enables studies that were not possible until recently. This is 
particularly true for astrometric data provided by SDSS. A good example is
the SDSS observations of asteroids: in only a few years, the SDSS has increased
the sample of asteroids with accurate colors by about two orders of magnitude,
and enabled detailed and robust studies of correlations between asteroid chemical 
and dynamical properties.

Another example where SDSS data were crucial is the construction of the Munn et al.
SDSS-POSS proper motion catalog. This resource is bound to make significant contributions
to our understanding of the Milky Way kinematics. The preliminary results discussed
here suggest that these kinematics may be significantly more complex than assumed
until now. 

\vskip 0.8cm
\acknowledgements 
Funding for the creation and distribution of the SDSS Archive has been provided by 
the Alfred P. Sloan Foundation, the Participating Institutions, the National Aeronautics 
and Space Administration, the National Science Foundation, the U.S. Department of Energy, 
the Japanese Monbukagakusho, and the Max Planck Society. The SDSS Web site is http://www.sdss.org/.

\end{document}